\documentclass[]{spie}  
\usepackage[]{graphicx}

\title{Antenna-Coupled TES Bolometer Arrays for CMB Polarimetry} 

\author{C.L. Kuo\supit{abc}, J.J. Bock\supit{cd}, J.A. Bonetti\supit{d}
J. Brevik\supit{c}, G. Chattopadthyay\supit{d}, P.K. Day\supit{d},
S. Golwala\supit{c}, M. Kenyon\supit{d}, A.E. Lange\supit{cd}, 
H.G. LeDuc\supit{d}, H. Nguyen\supit{d}, R.W. Ogburn\supit{c}, 
A. Orlando\supit{c}, A. Trangsrud\supit{c}, 
A. Turner\supit{d}, G. Wang\supit{ce} and J. Zmuidzinas\supit{cd} 
\skiplinehalf
\supit{a}Department of Physics, Stanford University, 382 Via Pueblo Mall, 
Stanford, CA 94305, USA;\\
\supit{b}Stanford Linear Accelerator Center, 2575 Sand Hill Rd., 
Menlo Park, CA 94025, USA;\\
\supit{c}California Institute of Technology, 1200 E California Blvd., 
Pasadena, CA 91125, USA;\\
\supit{d}Jet Propulsion Laboratory, 4800 Oak Grove Dr., 
Pasadena, CA 91109, USA;\\
\supit{e}Argonne National Laboratory, 9700 S. Cass Avenue, Argonne, IL 60439, 
USA
}

\authorinfo{For further information contact Chao-Lin Kuo, e-mail: 
clkuo@stanford.edu}

\begin{document} 
  \maketitle 

\begin{abstract}

We describe the design and performance of polarization selective 
antenna-coupled TES arrays that will be used in several upcoming Cosmic 
Microwave Background (CMB) experiments: SPIDER, BICEP-2/SPUD. The fully 
lithographic polarimeter arrays utilize planar phased-antennas for collimation 
(F/4 beam) and microstrip filters for band definition (25\% bandwidth). These 
devices demonstrate high optical efficiency, excellent beam shapes, and 
well-defined spectral bands. 
The dual-polarization antennas provide well-matched beams and low cross 
polarization response, both important for high-fidelity polarization 
measurements. These devices have so far been developed for the 100 GHz 
and 150 GHz bands, two premier millimeter-wave atmospheric windows for CMB 
observations. In the near future, the flexible microstrip-coupled architecture 
can provide photon noise-limited detection for the entire frequency range of 
the CMBPOL mission.
This paper is a summary of the progress we have made since the 2006 SPIE
meeting in Orlando, FL.

\end{abstract}



\section{Introduction}
\label{sect:intro}  

The primary science goals of CMB cosmology in the next decade are the 
degree-scale B-mode polarization 
induced by a gravitational wave background and the arcminute-scale B-mode 
induced by weak gravitational lensing from the large scale structures. 
The former will provide invaluable information on Inflation and early 
universe, while the latter offers a sensitive and complementary
probe of the dark energy and the neutrino mass. 
To achieve these challenging goals, the instruments will require a large 
number of sensitive mm-wave detectors, wide frequency coverage for astronomical 
foreground monitoring, and exquisite control 
of polarization systematics\cite{weiss05}. 

Bolometers can provide photon noise-limited sensitivity over a wide frequency 
range. In the past few years many groups have been working to improve the 
scalability of bolometric polarimeters over the existing feedhorn-coupled 
micromesh bolometers\cite{jones03}. 
These various efforts were surveyed in the NASA CMB Taskforce report in 2005
\cite{weiss05}. On the detector end, a very promising approach is the
microstrip-coupled superconducting transition edge sensors (TES), with 
microstrip inline filters to define science bands \cite{myers05,chuss06,kuo06}. 
The migration from semiconductor 
bolometers to TES bolometers enables the readout of thousands 
of pixels with moderate electronics complexity. This is achieved by
superconducting quantum interference device (SQUID) multiplexers 
\cite{dekorte03,lanting05}, now a mature technology. 
In addition, the thermally active components (the bolometers) do not scale 
with the wavelengths 
in a microstrip-coupled architecture, therefore the entire frequency range 
($\sim$ 30 GHz to 500 GHz) of interest in CMB science can be covered by the 
same technology. 


Because of these advantages, microstrip-coupled TES has been the 
technology of choice for the majority of the bolometer groups. However,
different groups approach the beam formation problem differently:
hemispherical silicon lenses \cite{myers05}, 
corrugated feed arrays from stacked metal platelets \cite{chuss06},
and metalized micromachined silicon frontend
are all being pursued.
The CMB polarimeter described in this paper is based on 
planar slot-antenna arrays \cite{kuo06}. The signal is summed coherently 
by a network of niobium microstrips to synthesize a collimated beam.
{\bf The key advantage for this approach 
is that the detector array is completely lithographic}. The only add-on piece 
is a thin, flat quartz wafer for anti-reflection coating. This 
greatly reduces the fabrication complexity and cost, and allows for very 
high focal-plane packing density. The monolithic arrays are immune to 
differential thermal contraction and misalignment. 

Co-locating dual-polarization detection is another key feature of the detectors 
described here. Bolometric receivers heavily rely on differencing 
to measure polarization. It is highly advantageous to measure orthogonal
polarizations simultaneously with beams that are as well-matched as possible. 
So far, the most successful bolometric polarization experiments all take 
advantage of co-locating dual-polarization bolometers 
\cite{boom_ee,yoon06,hinderks08}.

In SPIE 2006, we reported the performance of a series of prototype planar 
antenna-coupled TES detectors \cite{kuo06}. These first generation detectors 
provide good beam/band definition and demonstrate the feasibility of this 
technology. However their performance was not adequate for CMB applications.
Since then we have made a lot of progress in the design and fabrication. 
The major improvements are: (1) A better dual-polarization antenna design 
that has over 30\% bandwidth and intrinsically low cross-polarization and
instrumental polarization (beam mismatch); (2) A new inter-layer dielectric 
material (SiO$_2$) and deposition process (voltage-biased sputtering) that 
allows for better step coverage of the microstrips; and finally, (3) replacing
the contact masks with stepping lithography, greatly improving the overall 
precision and accuracy of the structures over the large array area.
After these improvements, the new detectors routinely achieve high 
polarization efficiency (97-99\%), high optical efficiency ($>$70\%), and
sufficient bandwidth (30\%). 
These characteristics are comparable or better than
the polarization sensitive detectors used in BICEP\cite{yoon06}, 
QUAD\cite{hinderks08}, and Planck. 

This paper is organized as follows. We first describe the architecture 
of a dual-polarization antenna-coupled bolometer in 
\S\ref{sec:components}. The fabrication, measured optical 
properties, and array performance are presented in \S\ref{sec:results}. 
We describe several CMB polarization experiments that will be using this 
technology and future development plans in \S\ref{sec:plans}.


\section{A Lithographic Polarimeter}\label{sec:components}

\begin{figure}[tbp]
\begin{center}  
\includegraphics[scale=0.7]{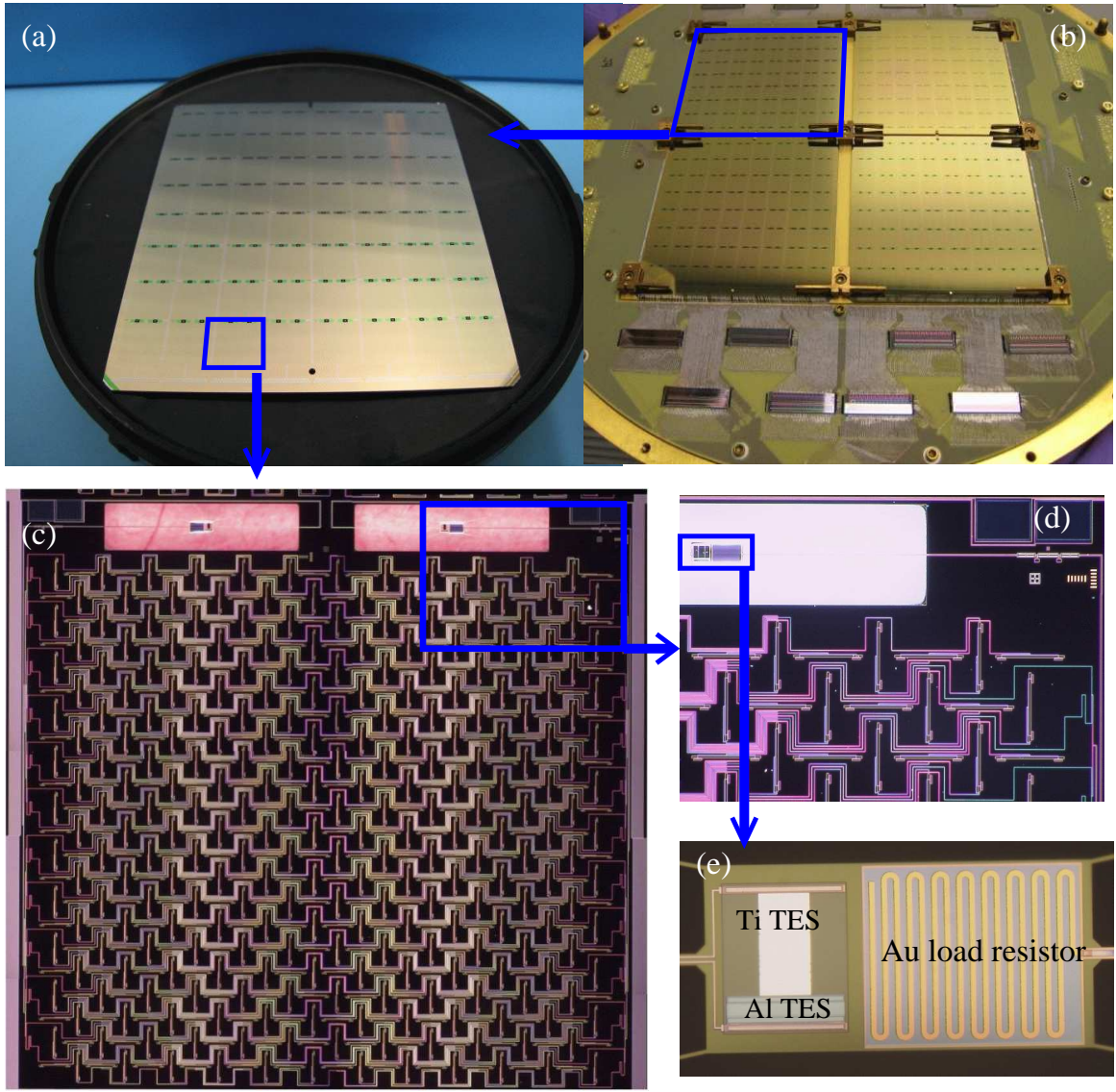}
\caption{This composite picture shows the different scales of the 
antenna-coupled TES polarimeter array. 
(a) An $8\times 8$ array of polarimeters fabricated on a 4-inch silicon
wafer. Each (small) square is a complete polarimeter unit. (b) Four such arrays 
are tiled together to make a focal plane. The number of polarimeter elements
depends on the frequency. At 145 GHz, there are 256 dual-polarization 
elements per focal plane, or 512 TES bolometers. Also visible are the 
32-element SQUID multiplexers and the ``Nyquist chips'' from NIST, and the 
PCB with superconducting aluminized traces (\S\ref{subsec:uniform}). 
(c) A polarimeter unit
consists of a pair of co-locating orthogonal antennas, the summing networks,
filters, and two TES bolometers, one for each linear polarization. 
The size of the 
polarimeter is $\sim 7.5$ mm at 145 GHz. (d) Details of 
the polarimeter. The antenna, summing tree, in-line microstrip filter
(near the upper right corner), and the TES (upper left) are shown. (e) 
Details of the thermally isolated TES bolometer. The meandering lossy 
microstrip is the termination resistor. The aluminum TES and titanium TES 
are connected in series for both laboratory-testing and science-observation 
loading conditions. 
}\label{fig:antenna_pic}
\end{center}
\end{figure}

Figure \ref{fig:antenna_pic} shows the overall structure of the focal plane 
arrays and the relative scales of various components. A polarimeter unit
consists of an array of slot antennas, summing networks, band-defining 
filters, and the TES bolometers, described below.

\subsection{Dual-polarization slot antennas}

\begin{figure}[tbp]
\begin{center}  
\includegraphics[scale=0.8]{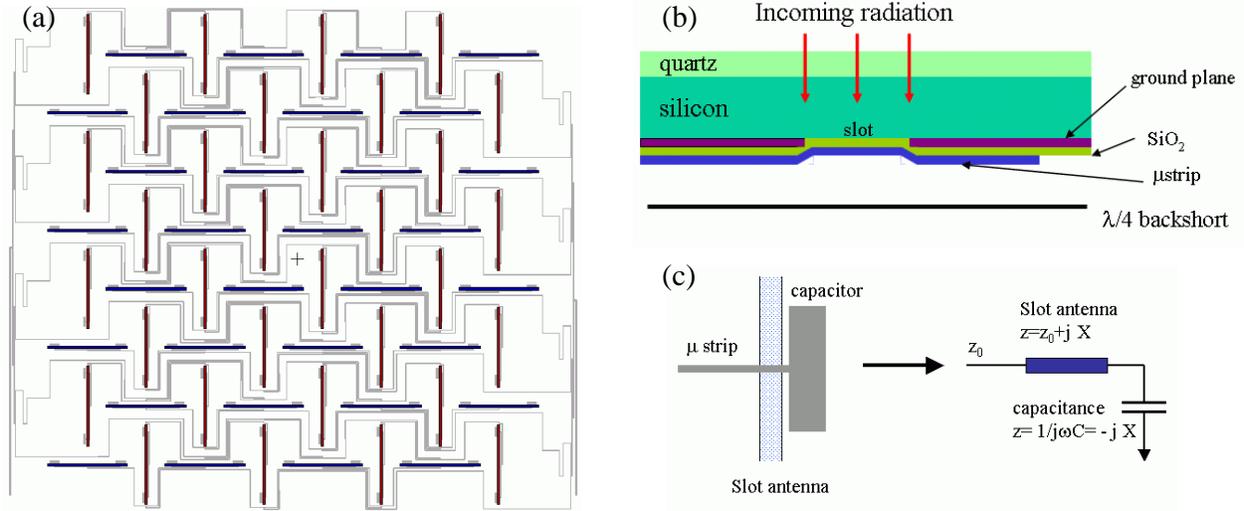}
\caption{(a). A demonstrative layout diagram of the dual-polarization slot 
antenna and the summing networks. Notice the $90^\circ$ rotational symmetry 
between two linear polarization with respect to the center. 
A simplified 4$\times$4 cell device is shown. The actual device consists of 
12$\times$12 or 10$\times$10 cells. 
(b) The cross section view of the slot antenna and microstrips. The radiation 
is coming through the AR-coating quartz layer from the clear silicon side,
exciting a voltage difference across the slots in the niobium ground plane.
The electrical field is inductively read out by a coupling capacitor. (c) The
coupling capacitor compensates the reactance produced by the offset-fed 
slot antenna to impedance-match the largely real impedance of the microstrip
lines.
}\label{fig:antenna}
\end{center}
\end{figure}

A collimated beam limits the radiation background onto the bolometer, 
and reduces side-lobes, stray light coupling, and susceptibility to cryogenic 
temperature fluctuations. Feed-coupled bolometers are especially 
advantageous compared to bare arrays under low background loading conditions 
at millimeter wave frequencies \cite{griffin02}. In an 
antenna-coupled detector, a planar array of slot antennas perform the function
of beam collimation. The signals coming from the sub-antennas are combined 
coherently by a microstrip summing network to form a beam. The beam width is
approximately given by $\lambda/d$, where $d$ is the linear dimension of the
antenna and $\lambda$ the wavelength in vacuum (Figure \ref{fig:antenna}a).

Long slots in a ground plane are intrinsically polarization sensitive since
microwave radiation tends to excite electric fields across the slots.
Another key motivation for choosing a slot architecture is that most of the 
substrate remains metalized, shielding the summing network from the 
incoming radiation. The cross section of the antenna structure is shown 
in Figure \ref{fig:antenna}(b).
As mentioned earlier, we seek a design that provides co-locating 
dual-polarization detection and enough bandwidth ($>$30\%). These requirements 
are not easy to meet simultaneously, because periodic slot structures are 
intrinsically narrow banded, and usually pose significant challenge for 
implementaing the summing network. 
The first generation dual-polarization detectors reported in our previous paper
\cite{kuo06} either produce excessive cross-polar response or span only 
15-20\% bandwidth. 

After several iterations we arrived at the design shown schematically 
in Figure \ref{fig:antenna}(a). 
The design has two sets of orthogonal slots, readout by two independent 
microstrip networks. 
Each slot is offset-fed to provide the appropriate impedance for the microstrip.
This is done symmetrically on both ends of the slot to ensure low 
cross-polarization response. 
The interleaved structure allows for longer slots and sufficient gaps 
to accommodate the summing networks.
The offset-feeding scheme produces significant 
reactance ($Z=37+15j \;\;\Omega$), which is compensated by an 
appropriate coupling capacitor 
(Figure \ref{fig:antenna} c) to match the largely real impedance of the 
microstrip feed lines. Method of moment and HFSS calculations show that this 
antenna has over 35\% bandwidth. 

Dual-polarization antennas make efficient use of the focal plane real 
estate, and more importantly, reduce polarization artifacts associated with 
differencing and pointing errors \cite{weiss05}. From Figure 
\ref{fig:antenna}(a) it is apparent that in our design the vertical and 
horizontal slots transform into each other after the device undergoes a 
$90^\circ$ rotation around the array center. In other words, the ``dipole'' 
components of beam mismatch vanishes \cite{hu2003,weiss05,hinderks08} by 
design. The detailed antenna parameters and impedance 
calculation will be published elsewhere \cite{day08}. 

\subsection{The summing networks}

The signals from the sub-antennas are combined in phase by a 
superconducting niobium (Nb)
microstrip network and transmitted to the TES. Two independent networks
are required, one for each linear polarization. 
The sidelobe response of the antenna is determined by how the
sub-antennas are excited by the summing network. If the sub-antennas are
arranged in a square grid pattern and fed uniformly, the radiation pattern 
will exhibit minor sidelobes (at -15dB) and a four-fold symmetry.
For reflector systems, it is desirable to minimize the sidelobe responses. 
This can be achieved by redesigning the feed network to taper the excitation 
pattern. In all current antenna designs, the sub-antennas are excited equally 
by the feed network. The minor sidelobes will be terminated at a cold stop
of the planned refractor systems. 

\begin{figure}[tbp]
\begin{center} 
\includegraphics[scale=0.75]{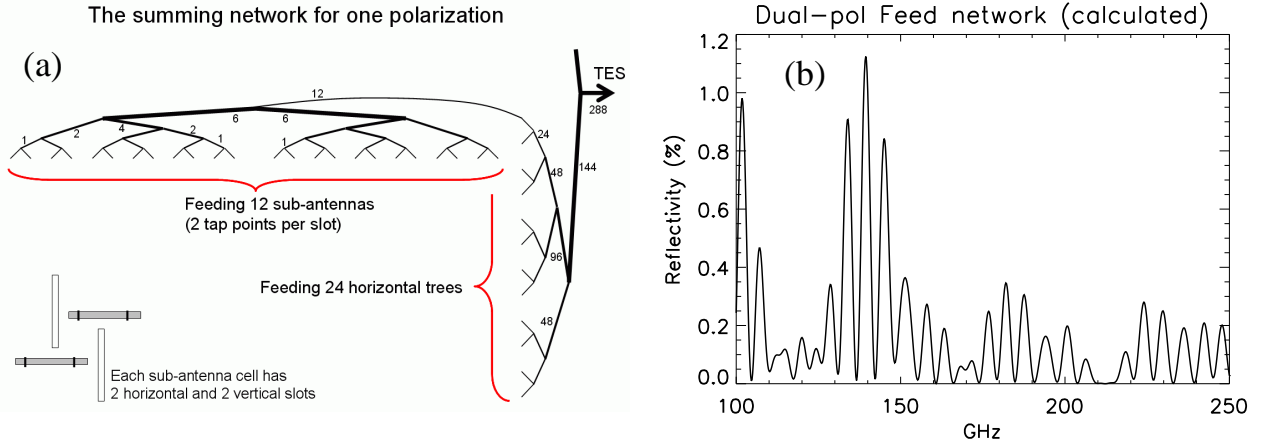}
\caption{{\em Left:} The schematic diagram of the feed network for one
of the two polarizations. An antenna consisting of $12\times 12$ cells is 
shown. The number next to each microstrip is the weight factor $w$, 
representing the number of upstream sub-antennas that are fed by this 
branch. 
{\em Right:} The calculated reflectivity of the summing network for a 
145 GHz antenna.
With proper tapering the reflectivity is restricted to less than $1\%$ over a
wide frequency range. The microstrip loss is neglected in the calculation. 
}\label{fig:tree}
\end{center}
\end{figure}

We use a combination of SuperMix library\cite{ward99} and SONNET 
simulation software to optimize the design of the summing network.
At the sub-antennas, the microstrip has to match the real part of the 
slot impedance. The signals from sub-antennas with the same polarization are 
combined by a network of T-junctions, as shown in Figure \ref{fig:tree}(a). 
These matched T-branches decrease the impedance downstream, since 
$1/Z_{joined}=1/Z_1+1/Z_2$. 
Tapered impedance transformers between T-junctions have to be used to 
maintain the downstream microstrips at manageable widths. 

For a given application, the geometry of the sub-antenna cell (the slot
length, width, gap, etc.) is determined by the desired operating impedance,
while the array size is fixed by the wavelength and 
the desired FHWM. These constraints prevent the use of simple binary trees 
and further complicates the design of the summing network. 
For $Z\sim 30-40 \Omega$ and FWHM $\sim 14^{\circ}$, 
the array format is usually in the $12\times 12$-cell range.
Finally, the full summing network, made of multi-level tapered 
microstrips and branches, has to be routed through the gaps between the slot 
sub-antennas and connected to the TES (Figure \ref{fig:antenna}a).

The seemingly daunting design challenge is greatly simplified by the following 
observation. We assign a weight factor $w$ to each microstrip branch, defined 
to be the number of sub-antennas from which the signal on the 
microstrip originated, as shown in Figure \ref{fig:tree}(a). For an 
impedance-matched summing network, the product of the local impedance $Z$ and 
the weight factor $Z\times w \equiv \tilde{Z}$ should only be a function of 
the electrical distance to the slots, and {\em independent} of the particular 
location of that microstrip branch. For a fixed total length and fixed end 
impedances, this reduces to a one-dimensional calculus of variation 
problem for ${\tilde Z}(x)$. The well-known solution to this problem is the 
Klopfenstein taper \cite{pozar}. 
For simplicity we adopt an exponential taper ${\tilde Z}(x)\propto \exp^{ax}$, 
and use piecewise linear microstrip transformer sections as an approximation. 
The reflectivity can be calculated by\cite{pozar} $\Gamma=\frac{1}{2}\int 
dx e^{-2\beta j x}\frac{d}{dx}\ln\left(\frac{{\tilde Z}}{z_0}\right)$. 
For our taper design
the reflected power is less than 1\% over a wide frequency range, see Figure 
\ref{fig:tree}(b). 
Since $w$ can be defined everywhere on the network, once a good design for 
${\tilde Z}(x)$ is found, the local impedance $Z$ on an arbitrary point on 
the network can be obtained by $Z={\tilde Z}/w$. The summing network is then 
laid out with the LEdit software. Following these procedures, we have corrected 
a design error in the feed network of a first generation antenna that has 
caused excessive reflections \cite{kuo06}. 


\subsection{The microstrip in-line filters}

In the planned CMB applications, the combination of absorptive, quasi-optical, 
and microstrip filters are used to reject out-of-band radiation. In particular,
the microstrip filters define both the upper and lower frequency cutoff of the 
science bands. The choice of the frequency bands in a CMB experiment is a 
trade-off between sensitivity, foreground information, detector readout 
counts, and optical performance, all under the constraint of a fixed focal 
plane area. Although $>$30\% bandwidth is possible with our antenna design, 
we have chosen a nominal 25\% fractional bandwidth for the bandpass filters, 
mainly because of the bandwidth limit of the half-wave plate in the optical 
train.  

 We have developed a lumped-element 3rd order bandpass Chebyshev 
$LC$ filter, consisting of CPW inductors and stub capacitors. This compact 
filter design does not include any ``vias'' (direct electrical contacts to the 
ground plane), and is fully compatible with photolithographic processes.
Because of their non-resonant nature, these filters do not have fundamental 
harmonic leaks. The bandgap frequency ($\sim $690 GHz) of niobium microstrips 
provides a natural high frequency cutoff for CMB experiments. In the past
3 years we have fabricated and tested more than a dozen such filters, with 
operating frequencies ranging between 90 and 250 GHz. Every tested filter
so far shows sharp low and high frequency cutoff, high in-band transmission,
and  very low high frequency leaks. The challenge has been to repeatedly and 
reliably produce filters with the correct center frequency to one- or two- 
percent level, and to control the detailed shape of the transmittance. 
The measured spectra of the first generation devices showed higher 
central frequencies then the calculated values \cite{kuo06}. The cause was 
traced to the fabrication uncertainty of the dielectric thickness. This issue 
has been resolved, by using a 6-inch SiO$_2$ target and better monitoring of 
the dielectric thickness during and after deposition. 
See \S\ref{subsec:optical} for the measured performance of the antennas 
and filters. 


\subsection{The TES bolometers} 

After the bandpass filter, the signal collected by the antenna is transmitted
through the Nb microstrip, and readout by a microstrip-coupled, thermally 
isolated bolometer on a micro-machined silicon nitride (SiN) island. 
The microstrip enters the thermally isolated patch via a suspended
SiN leg, and terminates in a meandering resistive microstrip 
(see Fig.\ref{fig:antenna_pic}e). 
A TES film deposited on the isolated region, and readout via superconducting Nb
leads, detects the heat from dissipation of electromagnetic energy in the 
resistor. The bolometer operates in the standard voltage bias configuration, 
which provides strong electrothermal feedback \cite{irwin95}. The principal 
benefits of this operating mode are linearity and rapid speed of response. 

The termination resistor is made of a meander of gold open-ended 
microstrip. The un-absorbed EM wave is reflected at the end of the microstrip, 
as a result the effective length of the resistive microstrip is twice the 
physical length (2.2 mm).
Since the characteristic impedance is largely determined by the geometry, 
the impedances of the superconducting and resistive microstrips are 
well-matched. The termination efficiency is calculated to be $99\%$ at 
100 GHz. The advantages of this design are wide bandwidth and immunity 
to variations in the thickness and the resistivity of the lossy film.
The same meandering microstrip can be used to terminate 30 GHz radiation
with an efficiency of $92\%$. A 3 mm long microstrip 
will absorb up to $97\%$ of the incoming power\cite{kuo06}.
Another significant advantage of microstrip-coupled bolometer geometry is
that the thermalization time constant is much shorter than the external 
(bolometer) time constant, providing excellent thermal efficiency. 

The TES for the sub-orbital CMB experiments 
will be based on elemental titanium (Ti). It is known that the $T_c$ of Ti films
is close to the bulk value and is fairly immune to thickness
variations. To facilitate 
laboratory testing under high optical loading conditions, we developed 
a dual-$T_c$ TES (0.46 K and 1.2 K) made from Ti and Al films (Figure 
\ref{fig:antenna_pic}e). 
We carry out dark electrical and thermal tests of 
a prototype SPIDER\cite{montroy06} bolometer in a dilution 
refrigerator using a SQUID sensor (not multiplexed). 
We find that when biased at the Ti transition the TES exhibits white noise 
between 0.1 and 10 Hz, with an 
electrical NEP of $\sim 2\times 10^{-17} W/\sqrt{Hz}$. The natural time 
constant is around 20 ms, and reduces to 1 ms with electrothermal feedback 
\cite{bonetti08}. The uniformity of the array parameters measured with a 
time-domain SQUID multiplexing system is reported in \S\ref{subsec:uniform}.


\section{Array Fabrication and Performance}\label{sec:results}

\subsection{Fabrication}\label{subsec:fab}

The antenna-coupled TES arrays are fabricated in the Microdevices Laboratory 
at JPL. The fabrication is accomplished using an 
assortment of lithography, deposition, and etching technologies.  The process 
is quite challenging and many technical hurdles were overcome to make the 
process reliable. The early problems we have encountered include TES wiring 
shorts by the Nb residue, pinholes in the dielectric layer, microstrips peeling 
off under stress, and poor metal-dielectric adhesion due to HCl contamination. 
In the past few months we have reached the point where arrays 
with good yields can be reliably produced. 
The current process requires no wet 
etches, and can be carried out by patterning only the front side of the wafer. 

\begin{table}[th]
\caption{The fabrication of various layers in an antenna-coupled TES 
array. } 
\label{tab:fab}
\begin{center}       
\begin{tabular}{|c|l|l|} 
\hline
\rule[-1ex]{0pt}{3.5ex} Material & Deposition & Patterning \\
\hline
\rule[-1ex]{0pt}{3.5ex}  Al& E-beam evap.& Lift-off\\
\hline
\rule[-1ex]{0pt}{3.5ex}  Ti& DC sputtering& ICP, Freon-12\\
\hline
\rule[-1ex]{0pt}{3.5ex}  SiO$_2$& RF sputtering& ICP, CHF$_3$/O$_2$\\
\hline
\rule[-1ex]{0pt}{3.5ex}  Nb& Magnetron sputtering& ICP, Cl$_3$/BCl$_3$\\
\hline
\rule[-1ex]{0pt}{3.5ex}  Au& E-beam evap.&Lift-off\\
\hline
\rule[-1ex]{0pt}{3.5ex}  SiN& Grown on Si& ICP, CHF$_3$/O$_2$\\
\hline
\rule[-1ex]{0pt}{3.5ex}  Si& NA& STS DRIE, XeF$_2$\\
\hline
\end{tabular}
\end{center}
\end{table} 

Since 2006, all lithography is performed with a Canon FPA 3000, deep 
ultra-violet, stepper mask aligner, allowing us to step and repeat 
many patterns on a 4-inch wafer. We found it necessary to use 6-inch 
Ti and SiO$_2$ targets in order to achieve adequate thickness uniformity.

We start with silicon wafers that have a 1 $\mu m$ thick layer 
of silicon nitride grown on the surface.
The initial step is creating the Al-Ti dual-$T_c$ TES. 
After the Ti etch, we immediately dip the wafers in water to wash away any HCl 
that may have formed from Cl residue and moisture in the air. 
We find it necessary to introduce SiO$_2$ passivation layers after the
creation of the Dual-Tc TES structures, and between the Al and Ti depositions. 
The electrical connection between the two TESs and with the Nb wiring layer 
is made with etched windows on the passivation SiO$_2$ layers.

The Nb ground plane is deposited by magnetron sputtering. Slots in the ground 
plane are patterned and ICP etched, then the wafers are again dipped in water. 
We then deposit the inter-layer dielectric (ILD) layer which separates the 
ground plane from a top Nb layer. The ILD is 285nm of SiO$_2$ and is 
deposited by biased RF sputtering. The biasing is necessary to insure nice, 
smooth contours at step edges. The thickness of this layer is vital to 
ensuring the microstrips and filters perform as desired.   

The uniformity results reported in 
\S\ref{subsec:uniform} came from an early engineering grade wafer
where a large fraction of the TESs are shorted by pinholes in the ILD. 
In order to solve this problem we create the 
ILD layer in two depositions.  After the first deposition, the sample is 
removed, polished, ultrasounded and O$_2$ plasma cleaned. The polishing 
dislodges any debris that may be loosely bound to the SiO$_2$, opening up 
holes.  The second SiO$_2$ deposition fills in these holes.  
After the ILD has been deposited it is patterned and etched to create 
vias for connecting the TES's to the top Nb layer. 
The top layer Nb is deposited, patterned, and etched with the same system, 
chamber, and parameters as the Nb ground plane layer.  The top Nb layer 
contains the summing networks, TES DC wiring, and filters. 
The next step is creating the termination resistor (Au meander),  
which completes the electrical aspects of the TES array. 

The final step is defining the thermally isolated membranes,  
perhaps the most challenging part of the fabrication process. 
The membranes are 150$\times$300 
$\mu m^2$ rectangles and are suspended and isolated by four, 3-$\mu m$ wide 
legs, and two, 9-$\mu m$ wide legs. The DC wiring and microstrip lines rest 
on the wider legs. The first step in defining the membranes is to etch the SiN 
in unwanted areas.  The patterning for this is done with a 4 $\mu m$-thick
photoresist. This etch exposes bare silicon in regions except those where the 
membrane and legs will be.  After the SiN etch, a very thick 
(~10 $\mu m$) photoresist is patterned to define the membrane, and membrane 
legs. A thick resist is necessary because the exposed bare silicon 
is etched with an STS deep trench etcher, which cuts completely through the 
500 $\mu m$ thick wafer in ~3 hours.   
The resist is not removed after the deep trench etching and is still thick 
enough to be used in the final step, a XeF$_2$ gas etch which undercuts the 
silicon underneath the membrane and legs to thermally isolate the bolometers. 
Table \ref{tab:fab} summarizes the fabrication methods for various layers.

\begin{figure}[h]
\begin{center}  
\includegraphics[scale=0.8]{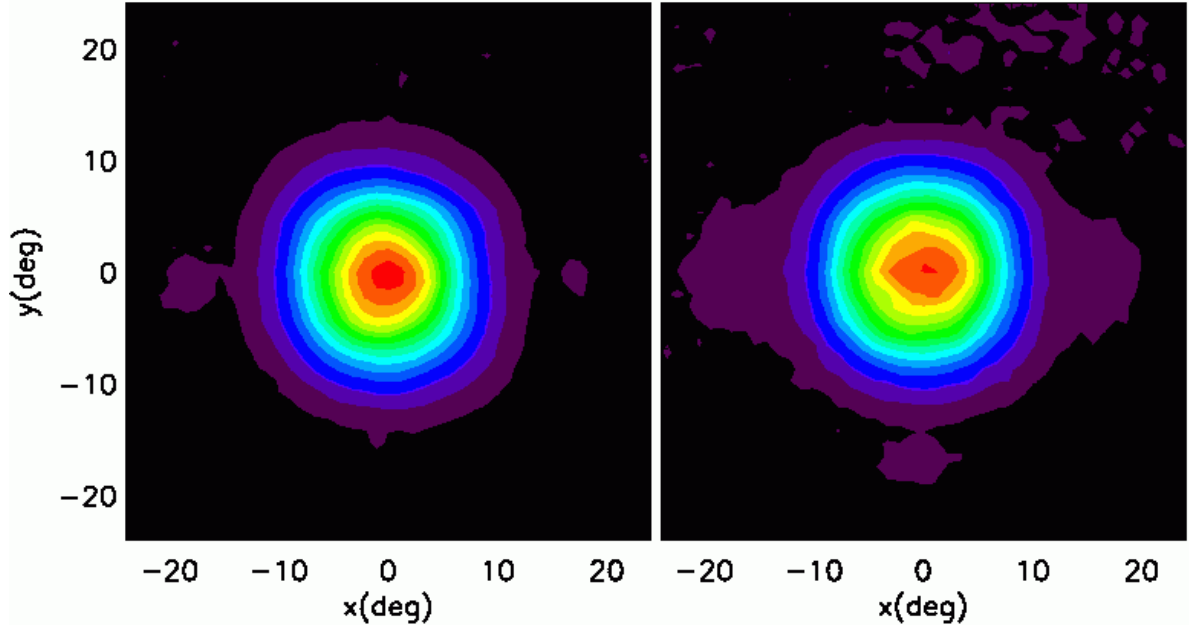}
\caption{The two co-polar radiation patterns for orthogonal polarizations 
of a filtered, planar antenna-coupled TES polarimeter. These are measured with 
a chopped broadband thermal source therefore the beams are averaged over the
25\% frequency band of the microstrip filter.
}\label{beams}
\end{center}
\end{figure}

\subsection{Optical testing and results}\label{subsec:optical}

At the time of writing the full receiver focal plane is still 
under construction and is not ready for array-level optical testing. 
We obtained optical data from individual detectors diced 
from an array. The detectors are tested in an optical testbed equipped 
with a closed-cycle $^3$He refrigerator. This is the same testbed we used to 
make the measurements reported in the previous SPIE paper \cite{kuo06}.
There are however some modifications to the system. First of all, we have 
reconfigured the optical train to eliminate metal mesh filters at oblique 
angles, which may have contributed to the cross polarization signal 
seen earlier. In the current setup, we use thick (1.5 cm), 
PTFE filters \cite{yoon06} at 77K and 4K for thermal filtering. 
These filters are mounted face-on, and anti-reflection coated to minimize 
polarization artifacts.
For some of the optical tests, we use an additional 2mm-thick carbon-loaded 
PTFE sheet to reduce in-band millimeter wave loading on the detectors. 
A $\lambda/4$ quartz plate is glued (GE varnish)
onto the silicon entry surface as the anti-reflection coating. In these tests 
the detectors are mounted in a copper box  with no $\lambda/4$ backshorts: the
summing tree side of the detector is facing a piece of eccorsorb absorber. 

The detectors are mounted in a superconducting niobium shield (with a 4.4-cm 
opening) during the measurements to guard the TES and the SQUID read-out 
against magnetic interference. For Al-Ti dual $T_c$ TES, the bolometers 
are biased at the Al transition. Some earlier devices have Mo-Au bi-layer
films with $T_c\sim$0.7K. 

The beam pattern (angular response) of the antenna-coupled TES is measured
with an optically modulated thermal source on a moving X-Y stage. 
Figure \ref{beams} show the measured co-polar radiation patterns for a detector
pair. The beams are symmetrical, and well-matched (to $\sim \%$ level).
Four minor sidelobes are visible, as predicted by physical optics models.
In a refractive receiver (such as BICEP-2 and SPIDER), the sidelobes are 
terminated at the cold Lyot stop. 
The illumination of the pupil, and consequently the beam on 
the sky, is then determined by the mainlobes of the antenna radiation pattern. 
The FWHM$_x$, FWHM$_y$ for the beams in Figure \ref{beams} are 
$(12.8^{\circ},13.1^\circ)$ and $(13.6^\circ,13.3^\circ)$, 
respectively. The center frequency for this earlier detector is 165 GHz.
After scaled to the nominal frequency of 145 GHz, the FWHM should be around 
$15^\circ$. 
Note 
that the dipole components of the differenced beam correspond to phase errors 
at the feed level and are not measurable with a thermal source.
In the planned experiments, half-wave plates will be incorporated in the 
optics to reduce systematics associated with beam mismatch. 

\begin{figure}[tbp]
\begin{center} 
\includegraphics[scale=0.7]{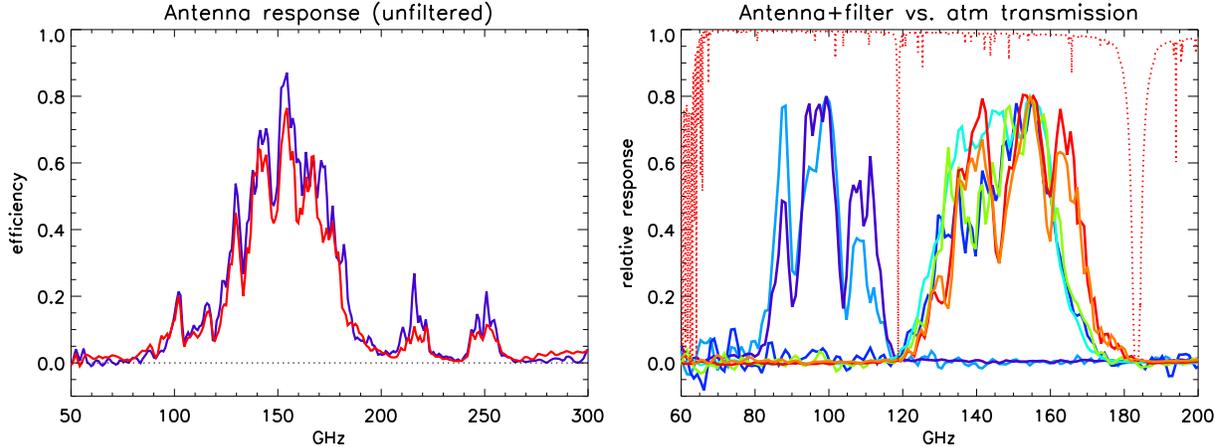}
\caption{
{\em Left} The spectral response of a pair of dual-polarization antennas 
without additional microstrip filters. These spectra are normalized using a 
temperature controlled blackbody source. The bare antennas show high efficiency
and over 30\% bandwidth.
{\em Right} The measured spectra for several filtered devices. The filter 
bands are within a few percent of the design values and are well-matched to the
atmospheric windows. For clarity these spectra are rescaled such that the peak
is at 80\%. 
}\label{fts1}
\end{center}
\end{figure}

The end-to-end spectra of the devices are measured with a Fourier transform 
spectrometer (FTS). The normalized spectra for a pair of unfiltered dual-
polarization antennas are plotted in the left panel (Figure \ref{fts1}a). We
assume single-moded throughputs ($A\Omega$), and correct for the effects of a
Mylar beam-splitter when plotting the spectra.
The observed center frequency agrees well with the design values (145GHz). 
The bandwidth is sufficient, but less than the predicted 39\%, perhaps as a
result of optical interference in the testbed, and to some degree, imperfect
fabrication. The features at high frequency are caused by surface wave modes 
in the silicon wafer \cite{day08}.
The FTS spectra in Figure \ref{fts1}(a) are normalized by measurements in a 
separate experiment, where the detectors are illuminated with a known optical 
power from a warm blackbody source (4.2K - 15K) \cite{kuo06}. Even without 
a $\lambda/4$ backshort, the end-to-end optical efficiency is over 70\% at the
peak, higher than most micromesh semiconductor bolometers
\cite{boom_ee,yoon06,hinderks08}. 

Figure \ref{fts1}(b) shows several representative spectra of the full 
antenna$+$filter devices. The plotted spectra are not absolutely normalized, 
but
the optical efficiency for a selected subset of the detectors are confirmed to
be high as well. The design center frequencies are 96 GHz and 145 GHz,
corresponding to two important atmospheric windows at mm-wavelengths. 
Also plotted is the atmospheric transparency at ballooning altitudes. The 
filter 
bands are within a few percent of the design values and are well-matched to the
atmospheric windows. There is evidence for optical interference in the
testbed and fabrication defects, which might have caused the observed structures
and variations in the transmittance curves.
We believe the shifts in the center frequencies between
different fabrication runs are entirely due to variations in the dielectric
thickness (SiO$_2$).

We used a free-standing wire grid polarizer\footnote{MicroTech G-40L.} to 
quantify the cross-polar response, also known as the polarization leakage. 
According to the manufacturer, the polarization leakage of the wire grid 
is between $10^{-4}-10^{-3}$ for the frequency range of $100-200$GHz . 
Therefore the measured peak cross-polar response, 1\% to 3\% is entirely 
due to the devices. 
This level of polarization leakage is comparable or better than the 
semiconductor-based feedhorn coupled polarization sensitive detectors
used in BICEP, QUAD, and Planck. The angular distribution of the 
cross-polar response has not been measured. 

The small cross-polar response, which varies between detectors, is most likely 
caused by unseen shorts, breaks, or other defects in the summing networks. 
This should be improved with better controlled fabrication procedures. 
As an example, the optically measured detectors are fabricated with a single 
ILD deposition (\S\ref{subsec:fab}) and might contain pinholes. A two-step 
ILD process has become part of the standard procedure and should improve the 
reliability of the summing networks. 

\subsection{Dielectric/microstrip testing}\label{subsec:sio2}

The performances of the summing networks and the filters 
depend on correct modeling of the dielectric and microstrip behavior.
The most important parameters are the dielectric constant and loss tangent 
of the ILD (SiO$_2$), and the kinetic inductance of the Nb microstrips. 

\begin{figure}[th]
\begin{center}  
\includegraphics[scale=0.8]{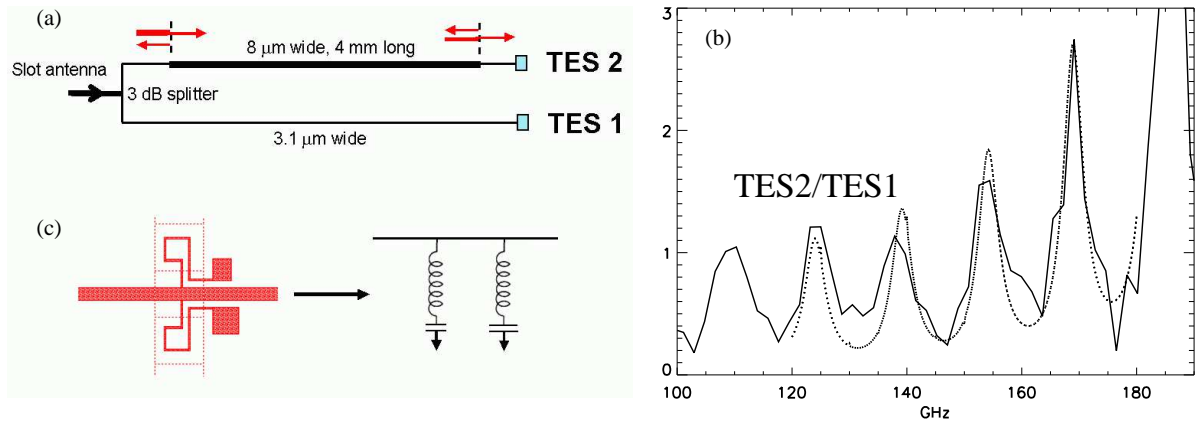}
\caption{(a). A schematic diagram of the SiO$_2$ test device for measuring the
phase velocity. The impedance discontinuities cause reflection,
creating interference fringes in the frequency spectra. 
(b). The ratio of the 
signals at TES2 versus TES1, with arbitrary normalization. The solid line is 
from the FTS measurements, and the dashed line is calculated using the SONNET 
software. The fringes are used to determine the phase velocity of the wide 
microstrip section in (a). (c). Another test device uses notch filters to 
measure the properties of SiO$_2$ and superconducting Nb. 
}\label{fig:sio2}
\end{center}
\end{figure}

We developed a series of TES devices aiming at measuring these properties.
In the first test device, 2 TES are coupled to a common single polarization 
slot antenna through a 3-dB power divider. One of the outputs of the power 
divider is connected to a reference TES (TES-1) through an impedance-matching 
microstrip. The other output is connected to TES-2 through a microstrip that 
has a wider section, 
see Figure \ref{fig:sio2}a. The reflection from the impedance discontinuities 
generates interference. 
We use the SONNET software to calculate the expected signal at TES-1 and TES-2,
and compare the results with the FTS measurements. The reflected wave can 
travel to TES-1, but the ratio of the signals in the 2 TESs exhibits prominent 
interference fringes. The frequency spacing of the fringes is a 
direct measurement of the phase velocity of the wide microstrip section, 
or equivalently, the imaginary part of the propagation constant $\beta$
(Figure \ref{fig:sio2}b). 
From the fringes we find that $\epsilon$ and $L_s$ fall along
the line $\epsilon=4.0-17.0\times(L_s-0.0935)$, where $L_s$ is measured in 
$pH$ per square. This result is quite insensitive to the ILD thickness of 
test device, since $\beta\propto \sqrt{LC}$, and thickness variation tends to 
cancel since it drives $L$ and $C$ in the opposite directions. 
The transmittance of the microstrip filter is used to break the degeneracy 
between $\epsilon$ and $L_s$, because in our filter design the inductance 
is dominated by geometrical inductance. Comparing SONNET simulations of the 
filters and the FTS measurements, we determined that 
$(\epsilon,L_s)=(3.9,0.10)$. 

A second test device consists of a pair of notch filters 
(Figure \ref{fig:sio2} c). Convincing notch frequencies are observed at 
133 GHz and 167 GHz, which agree to within 5\% of the SONNET simulations. 
However, through its dependence on $C$, this measurement is sensitive to the 
thickness variations of the ILD in the test device. We have decided to rely on 
the phase velocity measurements and the band-pass filters to derive the 
parameters, and use the notch filters only as a consistency check.

The third test device was developed to measure the loss tangent of the 
dielectric material SiO$_2$. The design is similar to the test device for
the phase-velocity measurements, with the wide microstrip section in Figure
\ref{fig:sio2}(a) replaced by a very long, meandering Nb microstrip (3.1
$\mu m \times$ 11.4 cm). The device is illuminated with a chopped thermal 
source, and the signals at TES2 and TES1 are compared. After correcting for 
the responsivity we obtain an attenuation constant $\alpha=4.6\times 10^{-2}$
cm$^{-1}$, corresponding to a dielectric loss tangent of 
$\tan \delta = 6\times 10^{-4}$
at 145 GHz. This translates into a negligible loss contribution from the
feed network in an antenna-coupled bolometer, which typically has 2-3 cm
of microstrip lines from the slots to the TES. This low loss is consistent with 
the measured high optical efficiency.

It is worth noting that the loss measurements are carried out at 
sub-Kelvin temperatures, and the modulated signal is small. This is important
because the dielectric loss is a strong function of temperature and excitation 
power \cite{loss77}. The loss tangent quoted above is representative of what we 
expect in a CMB experiment. 

\subsection{Array uniformity}\label{subsec:uniform}

We have conducted dark measurements of an antenna-couple TES array 
with the full SQUID multiplexing system.  
The TESs of the array are read out by 4 NIST SQUID multiplexer (MUX) chips,
 each MUX chip reading out 32 TESs.  Additionally, 
4 NIST ``Nyquist'' (NYQ) chips are used in conjunction with the MUX chips to 
roll off high frequency noise. The shunt(bias) resistors are fabricated on the 
NYQ chips. 
We employ time-domain multiplexing\cite{dekorte03} to read out the 
array in a 4 column by 32 row format, using the Multi-Channel Electronics 
(MCE) system developed at University of British Columbia.  The columns are 
defined by the 4 sets of NYQ/MUX chips, while the rows 
are defined by the 32 first stage SQUIDS each inductively coupled to a TES.  
The 4 columns (MUX chips) share the same sets of row addressing (RA) lines.
And we have one pair of bias lines per NYQ chip, whereby the 
32 TESs of a given column are biased in series.   

The 128-element array tile is mounted on a gold-plated copper stage, 
cooled to a base temperature of 260mK by a 3-stage (3He/4He/3He) Helium 
refrigerator.  It is held to the stage by beryllium-copper spring clips fixed 
near the corners of the array, with a thin layer of plastic shim between.
Also attached to the 260mK stage is a multi-layer printed circuit board, used 
as the attachment point for the NYQ and MUX chips.  The PCB also
provides routing for the control wiring for the NYQ/MUX chips and the signal 
lines that connect them to the TES array.  The TES signal lines are 
superconducting; formed on the top layer of the PCB by aluminum sputtered onto 
copper lines.  The NYQ/MUX chip control wiring is distributed within the 
multiple layers of the PCB, such that the signal and return lines of the row 
addressing lines overlap to reduce pickup loops.  Connection between control 
and signal wiring and the array and NYQ/MUX chips is made using 0.001'' 
aluminum wirebonds.
The NYQ/MUX chips are mounted on the PCB using an intermediary alumina 
carrier, which has been selected due to the similar coefficients of thermal 
expansion of alumina and silicon.  The carriers are first attached to copper 
pads on the PCB using indium solder.  The NYQ and MUX chips are expoxied to 
the carriers side-by-side. 
The array mounting/wiring scheme developed for this single array testbed 
remains unchanged in the 4-tile focal plane of the planned CMB experiments.
Figure \ref{fig:antenna_pic}(b) shows the mounting of 4 arrays and NYQ/MUX 
chips, the PCB with aluminized traces, and the Al wire bonds.

The array tested is a SPIDER engineering grade array, 
with a designed TES transition of 450mK, and a thermal conductance of 
$G_{450} = 19$ pW/K. Shortly after the array was mounted we discovered that 
many of the TESs suffer from the ILD pinholes and are shorted to the 
ground plane and the chassis. In addition, there is a first-stage SQUID 
failure and some wire bonding issues. The total number of operational TESs
is 60 out of 128. The yield will greatly increase in the future, since 
we have identified and improved the ILD pinhole problem and the 
SQUID/wire bonding quality should 
improve with more experience. The measured mean and RMS of the TES
parameters are summarized in Table \ref{tab:uniform}.

\begin{table}[th]
\caption{The measured mean and RMS of the TES parameters derived from 
measurements of 60 multiplexed TESs.}
\label{tab:uniform}
\begin{center}       
\begin{tabular}{|c|l|l|l|} 
\hline
\rule[-1ex]{0pt}{3.5ex} Parameter &  Mean & RMS & Target \\
\hline
\rule[-1ex]{0pt}{3.5ex} $T_c$ (mK) & 466 & 3.9 & 450 \\
\hline
\rule[-1ex]{0pt}{3.5ex} R$_n$ (m$\Omega$) & 61.9 & 10.8 & 60 \\
\hline
\rule[-1ex]{0pt}{3.5ex} G$_{450}$ (pW/K) & 13.6 & 1.67 & 19 \\
\hline
\end{tabular}
\end{center}
\end{table} 

\begin{figure}[th]
\begin{center}  
\includegraphics[scale=0.5]{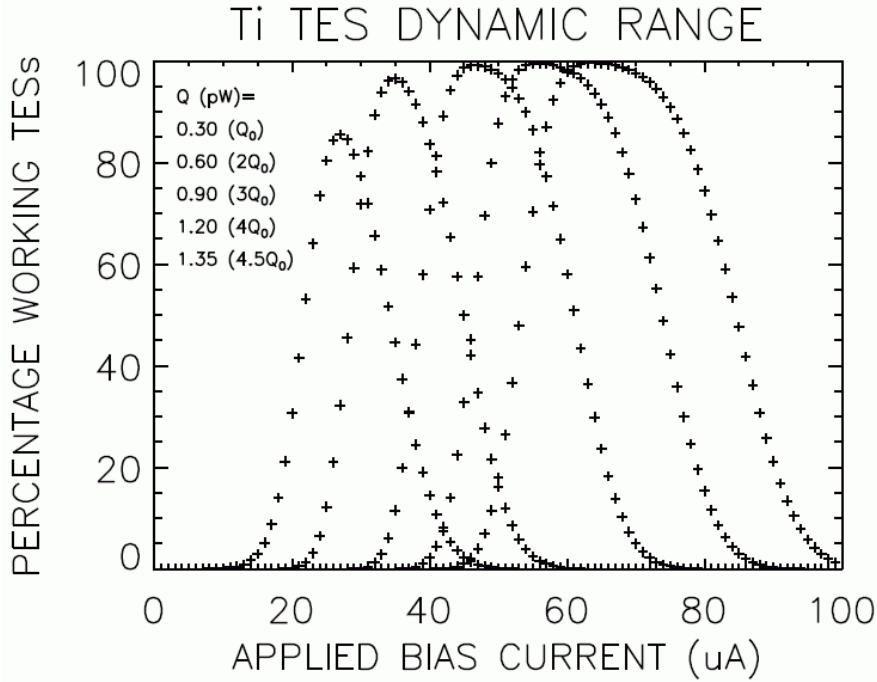}
\caption{Predicted SPIDER array yields, as a function of bias and optical 
loading. This is obtained from Monte Carlo simulations, using the measured
scatter in $(T_c,R_n,G)$ (Table \ref{tab:uniform}). The projected 
loading $Q_0$ is 0.3pW at 145 GHz, but the array can tolerate 4 times the 
nominal loading and still have 95\% of the TESs biased in the electrothermal 
feedback mode.
}\label{fig:uniform}
\end{center}
\end{figure}

Given these mean values and their spreads, a determination can be made of the 
number of functional TESs a system will have under a range of optical loading,
$Q$.  We used a Monte Carlo simulation to calculate the effect of the mean and 
spread of the values detailed above in the power balance equation for a TES.
\begin{equation}
Q\sim \frac{G_0}{(\beta+1)T_b^\beta}
(T_c^{\beta+1}-T_b^{\beta+1})-\frac{V_b^2}{R},
\end{equation}  
where $Q$ is the optical power, $T_b$, $T_c$ are the bath temperature 
and the transition temperature, and $V_b$ the bias voltage. 
$G=G_0(T/T_b)^{\beta}$ defines the temperature dependent thermal conductance of 
the bolometer. The equation was solved for the TES operating resistance, and 
a TES was considered operational if this value was between 0.3 and 0.95 times 
the normal resistance for a given applied TES bias.  
The fraction of operational detectors was calculated for a range of applied TES
 biases, which yields the optimal bias point for a given $Q$.  The $Q$ was then 
stepped through 0.3, 0.6, 0.9, 1.2 and 1.35 pW, and the family of curves was 
plotted in Figure \ref{fig:uniform}.  The figure depicts the dynamic range 
of the TES array given varying optical loading conditions.  The optical 
loading can increase from the nominal value of 0.3pW to 1.35pW with a loss of 
roughly 15\% of the TESs after properly adjusting the bias. 

\section{Future Plans}\label{sec:plans}

So far we have successfully demonstrated 145 GHz and  96 GHz devices, and have 
obtained a design for 225 GHz. We will continue to extend the frequency coverage
of this technology to both high and low frequencies, and to improve fabrication
and uniformity of the TES arrays. Other direction of detector development 
may be a detailed study of the dielectric/microstrip properties. Such 
information can be used to improve the design of the summing network and on chip
microstrip hybrids that can be used for simultaneous $QU$ detection. The summing
networks can be tapered to lower the sidelobes, or phase lagged to allow for
non-telecentric optical systems. 

These detectors will first be used in the upgrade of the BICEP experiment
\cite{yoon06}, ``BICEP-2'', and SPIDER, a balloon-borne experiment 
\cite{montroy06}.
BICEP-2 and SPIDER share common design concepts, including large throughput 
cold refractive telescopes that produce extremely low cross-polarization,
instrument polarization, instrument loading, and beam ellipticity. 
The 30 cm compact optics enables detailed pre-flight/deployment 
characterizations, and provides an angular resolution of $\sim 1^\circ$ suitable
 for primordial B-mode polarization.

Despite these similarities, the two experiments are based on very 
different observing strategies.
SPIDER is targeting very large angular scale CMB polarization, including the 
re-ionization bump at $\ell\sim 8$. It will survey 50\% of the sky.
To facilitate Galactic foreground removal on large angular scales, 
SPIDER payload consists of 6 telescopes, covering a wide 
frequency range (80-275 GHz) with $\sim$ 2000 antenna-coupled TES bolometers.
On the other hand, BICEP-2 and its successor SPUD will be observing from 
South Pole mainly through the 2mm and 3mm atmospheric windows. With the long 
integration 
time available from the ground, it will go extremely deep on 1-2\% of the 
sky that has minimum astronomical foregrounds.  

Both experiments will be important scientific pathfinders for CMBPol, a 
comprehensive NASA satellite to study the CMB polarization~\cite{weiss05}. 
In addition, many technical aspects discussed in this paper will be 
thoroughly tested. State-of-the-art 
photolithography can now reliably produce silicon nitride legs with
extremely high aspect ratio. The NEP required by 
a space-borne CMB mission is now routinely achieved in the laboratory
with appropriate time constant, using TES materials that have lower 
$T_c$ such as Mo/Cu or Mo-Au bi-layers. To produce bi-layers with highly 
controlled $T_c$ poses some fabrication challenge, but is definitely 
possible. The current SQUID MUX can multiplex up to 32 
channels~\cite{dekorte03}. In the future, the microwave frequency domain 
SQUID MUX ~\cite{irwin04} with reflectometer readout might be able to 
multiplex hundreds or even thousands of detectors, with electronics at a 
much lower cost. We believe the antenna-coupled TES detector technology is
a strong candidate for CMBPol. 

\acknowledgments     

The authors acknowledge support from the JPL Research and Technology
Development program, a NASA/APRA grant ``Antenna-Coupled TES
Bolometer Array for CMB Polarimetry'' to J. Bock, and the
Gordon and Betty Moore foundation.



\bibliography{tes}   

\begin{thebibliography}{10}

\bibitem{weiss05}
J.~{Bock}, S.~{Church}, M.~{Devlin}, G.~{Hinshaw}, A.~{Lange}, A.~{Lee},
  L.~{Page}, B.~{Partridge}, J.~{Ruhl}, M.~{Tegmark}, P.~{Timbie}, R.~{Weiss},
  B.~{Winstein}, and M.~{Zaldarriaga}, ``{Task Force on Cosmic Microwave
  Background Research},'' {\em astro-ph}~{\bf 0604101}, Apr.~2006.

\bibitem{jones03}
W.~C. {Jones et al.}, ``{A Polarization Sensitive Bolometric Receiver for
  Observations of the Cosmic Microwave Background},'' in {\em Millimeter and
  Submillimeter Detectors for Astronomy. Proceedings of the SPIE, Volume 4855},
   T.~G. {Phillips} and J.~{Zmuidzinas}, eds., pp.~227--238, Feb.~2003.

\bibitem{myers05}
M.~J. {Myers et al.}, ``{An antenna-coupled bolometer with an integrated
  microstrip bandpass filter},'' {\em Applied Physics Letters}~{\bf 86},
  pp.~4103--+, Mar.~2005.

\bibitem{chuss06}
D.~{Chuss et al.}, ``{The Primordial Anisotropy Polarization Pathfinder Array
  (PAPPA): Instrument Overview And Status},'' in {\em Proceedings of the SPIE},
   2006.

\bibitem{kuo06}
C.~L. {Kuo et al.}, ``{Antenna-coupled TES bolometers for CMB polarimetry},''
  in {\em Millimeter and Submillimeter Detectors and Instrumentation for
  Astronomy III. Edited by Zmuidzinas, Jonas; Holland, Wayne S.; Withington,
  Stafford; Duncan, William D.. Proceedings of the SPIE, Volume 6275, pp.
  62751M (2006).},  {\em Presented at the Society of Photo-Optical
  Instrumentation Engineers (SPIE) Conference}~{\bf 6275, astro-ph/0606366},
  July~2006.

\bibitem{dekorte03}
P.~{de Korte et al.}, ``{Time-division superconducting quantum interference
  device multiplexer for transition-edge sensors},'' {\em Review of Scientific
  Instruments}~{\bf 74}, pp.~3807--3815, Aug.~2003.

\bibitem{lanting05}
T.~M. {Lanting et al.}, ``{Frequency-domain multiplexed readout of
  transition-edge sensor arrays with a superconducting quantum interference
  device},'' {\em Applied Physics Letters}~{\bf 86}, pp.~2511--+, Mar.~2005.

\bibitem{boom_ee}
T.~E. {Montroy et al.}, ``{A Measurement of the CMB EE Spectrum from the 2003
  Flight of BOOMERANG},'' {\em Astrophysical Journal}~{\bf 647}, pp.~813--822,
  Aug.~2006.

\bibitem{yoon06}
K.~W. {Yoon et al.}, ``{The Robinson Gravitational Wave Background Telescope
  (BICEP): a bolometric large angular scale CMB polarimeter},'' in {\em
  Millimeter and Submillimeter Detectors and Instrumentation for Astronomy III.
  Edited by Zmuidzinas, Jonas; Holland, Wayne S.; Withington, Stafford; Duncan,
  William D.. Proceedings of the SPIE, Volume 6275, pp. 62751K (2006).},  {\em
  Presented at the Society of Photo-Optical Instrumentation Engineers (SPIE)
  Conference}~{\bf 6275, astro-ph 0606278}, July~2006.

\bibitem{hinderks08}
{QUaD collaboration: J.~Hinderks et al.}, ``{QUaD: A High-Resolution Cosmic
  Microwave Background Polarimeter},'' {\em astro-ph}~{\bf 08051990}, May~2008.

\bibitem{griffin02}
M.~J. {Griffin}, J.~J. {Bock}, and W.~K. {Gear}, ``{Relative performance of
  filled and feedhorn-coupled focal-plane architectures},'' {\em Applied
  Optics}~{\bf 41}, pp.~6543--6554, Nov.~2002.

\bibitem{hu2003}
W.~{Hu}, M.~M. {Hedman}, and M.~{Zaldarriaga}, ``{Benchmark parameters for CMB
  polarization experiments},'' {\em Physical Review D}~{\bf 67}, pp.~043004--+,
  Feb.~2003.

\bibitem{day08}
P.~Day~et al. {\em In preparation} , 2008.

\bibitem{ward99}
J.~{Ward}, F.~{Rice}, G.~{Chattopadhyay}, and J.~{Zmuidzinas}, ``{SuperMix: A
  Flexible Software Library for High-Frequency Circuit Simulation, Including
  SIS Mixers and Superconducting Elements},'' in {\em Proceedings of Tenth
  International Symposium on Space Terahertz Technology},  pp.~268--281, 1999.

\bibitem{pozar}
D.~M. {Pozar}, {\em {Microwave Engineering, 3rd Ed.}}, John Wiley \& Sons,
  2005.

\bibitem{irwin95}
K.~D. {Irwin}, ``{An application of electrothermal feedback for high resolution
  cryogenic particle detection},'' {\em Applied Physics Letters}~{\bf 66},
  pp.~1998--2000, Apr.~1995.

\bibitem{montroy06}
T.~E. {Montroy et al.}, ``{SPIDER: a new balloon-borne experiment to measure
  CMB polarization on large angular scales},'' in {\em Ground-based and
  Airborne Telescopes. Edited by Stepp, Larry M.. Proceedings of the SPIE,
  Volume 6267, pp. 62670R (2006).},  {\em Presented at the Society of
  Photo-Optical Instrumentation Engineers (SPIE) Conference}~{\bf 6267},
  July~2006.

\bibitem{bonetti08}
J.~A. {Bonetti et al.}, ``{Electrical and Thermal Characterization of
  Membrane-Isolated, Antenna-Coupled, TES Bolometers for the SPIDER
  Experiment},'' {\em Journal of Low Temperature Physics}~{\bf 151},
  pp.~138--143, Apr.~2008.

\bibitem{loss77}
M.~{von Schickfus} and S.~{Hunklinger}, ``{Saturation of the dielectric
  absorption of vitreous silica at low temperatures},'' {\em Physics Letters
  A}~{\bf 64}, pp.~144--146, Nov.~1977.

\bibitem{irwin04}
K.~D. {Irwin} and K.~W. {Lehnert}, ``{Microwave SQUID multiplexer},'' {\em
  Applied Physics Letters}~{\bf 85}, pp.~2107--+, Sept.~2004.

\end{thebibliography}
\bibliographystyle{spiebib}   

\end{document}